\documentclass[aps,floatfix,twocolumn,superscriptaddress]{revtex4}
\usepackage{amsbsy}
\usepackage{amsmath}
\usepackage{epsfig}

\newcommand{\bra}[1]{\langle#1 |}
\newcommand{\ket}[1]{|#1 \rangle}

\newcommand{\average}[1]{\langle #1 \rangle}

\newcommand{\innerprod}[2]{\left \langle #1 | #2 \right\rangle}
\newcommand{\Tr}{\mathrm{Tr}}
\newcommand{\matBasis}[2]{\ket{#1}\bra{#2}}

\begin{document}

\title{Operator fidelity susceptibility,
  decoherence and quantum criticality} \author{Xiao-Ming Lu}
\affiliation{Zhejiang Institute of Modern Physics, Department of
  Physics, Zhejiang University, Hangzhou 310027, China} \author{Zhe
  Sun} \affiliation{Zhejiang Institute of Modern Physics, Department of
  Physics, Zhejiang University, Hangzhou 310027, China}
  \affiliation{Department of Physics, Hangzhou Normal University,
  Hangzhou 310036, China} \author{Xiaoguang Wang}
\email{xgwang@zimp.zju.edu.cn} \affiliation{Zhejiang Institute of
  Modern Physics, Department of Physics, Zhejiang University, Hangzhou
  310027, China} \author{Paolo Zanardi} \affiliation{Department of
  Physics and Astronomy, University of Southern California, Los
  Angeles, CA 90089-0484 (USA)}
\affiliation{Institute for Scientific Interchange, Viale
Settimio Severo 65, I-10133 Torino, Italy}

\begin{abstract}
  The extension of the notion of quantum fidelity from the state-space level to the operator one
  can be used to  study  environment-induced decoherence. State-dependent operator fidelity suceptibility (OFS),
  the leading order term  for slightly  different operator parameters, is shown to have a  nontrivial behavior when
  the environment is at  critical points.  Two different contributions to
  OFS are identified which have distinct physical origins and  temporal dependence.
  Exact results for the finite-temperature decoherence caused by a bath described by the Ising model in transverse field
  are obtained.
\end{abstract}
\maketitle

\section{Introduction}
Generically the interaction between a quantum system and its
environment results in decoherence, and may lead the system to
experience the so-called  quantum-classical transition \cite{Zurek}.
For this reason the decoherence process  is  regarded as
the main obstacle for the implementation of quantum information processing \cite{Nielsen}.
Generally speaking the  properties of the environment may strongly
affect its decohering capabilities
\cite{Karkuszewski,Cucchietti,HTQuan,Cucchietti2,Cormick}. This
implies that a quantum system can be regarded as a probe to extract
useful information about the coupled environment, e.g., the quantum
phase transitions (QPTs). Very  recently a nuclear magnetic
resonance (NMR) experiment of this kind has been performed
\cite{JZhang}. There a single qubit was used as a probe to detect
the (precursors of the) quantum critical point of the coupled
environment.

The relationship between the decoherence of a central spin and the
QPTs of the coupled environment can be established through the
notion of Loschmidt echo (LE) \cite{HTQuan}.  It is well known that
LE can be exploited to measure the stability of a quantum system to
perturbations \cite{Jalabert,Karkuszewski,Gorina}. In
Ref.~\cite{HTQuan}, a central spin coupled with an Ising spin chain
in a transverse field was considered; the authors found that the
decay of LE is enhanced by quantum criticality. The connection
between LE and quantum criticality has been further clarified in
Ref.~\cite{Rossini}. There the authors showed that the LE
enhancement holds true just for short decay times (gaussian regime).
A genuine signature of criticality, on the other hand, can be recovered
by studying the asymptotic (large time) behaviour of LE as function
of the environment coupling parameter \cite{Rossini}. More recently,
the averaged LE over all states on Hilbert space with a Haar
measure, called operator fidelity, was proposed to study quantum
criticality \cite{XWang}. This proposal is a direct extension to the
operator level of state-space quantum fidelity. This latter notion
recently attracted a lot of attention as a new tool to analyze
quantum criticality, both a zero-temperature and at finite
temperature \cite{Zanardi,HQZhou,Zanardi5,Zanardi4,WLYou,SChen}.

 Whereas in \cite{HTQuan} the initial state of the  spin chain was assumed to be the ground state,
in this paper we aim at  considering a more general situation: the
initial state of the environment is a mixture of the eigenstate of its
Hamiltonian. The Gibbs thermal state is special instance of this setup.
With this assumption the study of decoherence naturally leads to a state-dependent generalizations
of the  operator fidelity suggested in \cite{XWang}.

In this paper we will discuss the operator extensions of fidelity
on general grounds. In order to do that one has simply to notice that
finite-time quantum evolutions correspond to unitary operators that  themselves belong to an Hilbert space i.e., the Hilbert-Schmidt
one. It follows that some of the results obtained in the fidelity approach for quantum states lifts immediately to the operator level
we are now interested in.
We will show that the  corresponding operator fidelity susceptibility (OFS)
contains two different contributions.
These two terms arise from variation upon parameter change of energy levels and eigenstates respectively
and have very different temporal behaviour.
We will then get a general expression for OFS for
models with a factorization structure typical of quasi-free models. Finally, we will exploit OFS for studying
an environment described by an Ising chain with a transverse field that can be driven to quantum criticality.

This paper is organized as follows:
 In Sec.~\ref{sec:model}, we give a
general description of the models and the relations between
decoherence of the central system and the operator fidelity of two
time evolutions of the environment. In Sec.~\ref{sec:unitary}, we
give generalities about operator metric and fidelities and analyze
the general OFS. In Sec.~\ref{sec:fact},  we consider the models
with a factorization structure, and give a general expression for
the OFS, and then study the Ising model with a transverse field, get
the exact solution for the OFS. Sec.~\ref{sec:conclusion} contains
the conclusions and outlook.

\section{Decoherence and operator fidelity \label{sec:model}}
As mentioned in the introduction,  an important physical  motivation for the
operator fidelities  we are going to analyze in this paper is provided by the
decoherence process. To see this clearly, let us  consider a purely
dephasing coupling of a system with its environment. The interaction
Hamiltonian has  the form
\begin{equation}
  \label{eq:dp_model}
  H_{I}=\sum_{n}\matBasis{n}{n}\otimes B_{n}.
\end{equation}
The entire Hamiltonian is $H=H_0+H_I$, where $H_0$ consists of system
Hamiltonian $H_S=\sum_n E_n \matBasis{n}{n} $ and bath Hamiltonian $H_B$.
The $H_S$ eigenstates $\{\ket{n}\}$, play here the role of preferred pointer states
\cite{Zurek}.

With the initial state $\rho (0)=\matBasis{\psi (0)}{\psi(0)}\otimes
\rho _{B}$, the reduced density matrix of the system at time $t$ can
be written as
\begin{equation}
  \rho _{S}(t)=\sum_{n,m}c_{n}c_{m}^{\ast }e^{-i(E_n-E_m)t}\ket{n}\bra{m} \Tr_{B}[\rho_{B}V_{m}^{\dagger }(t)V_{n}(t)],
\end{equation}
where $c_n=\innerprod{n}{\psi(0)}$, $V_n(t)=\exp\left [ -it(H_B+B_n)\right ]$

Obviously, in the generic case $\rho_S$ evolves from initial pure states to mixed states.
 The decay of the off-diagonal elements of $\rho_S$ means
a reduction from a pure state to a classical mixture of the
preferred pointer states --- quantum-classical transitions. The
temporal behavior of the off-diagonal element is decided by two
factors, one is $c_nc_m^*$, just relating to initial state, and the
other $\Tr_{B}[\rho_{B}V_{m}^{\dagger }(t)V_{n}(t)]$ is unrelated
with the initial state of the system, reflecting the dephasing
effect induced by environment. The latter can be considered as a
fidelity for two operators $V_m$ and $V_n$.  which is defined by
\begin{equation}
  F_\rho(V_m,V_n)=\mid\langle V_m,V_n \rangle_\rho\mid
\end{equation}
where the inner product $\langle V_m,V_n \rangle_\rho = \Tr[\rho
V_m^\dagger V_n]$.  This fidelity can be obtained from the
density matrix of the central system and encodes information about the
 bath state.
Notice that one might choose $B_n=\lambda_n B;$
in this case different $V_n$ simply correspond to different values
of the coupling strength $\lambda$ in front of the "perturbation"
$B.$ This is the scenario we mostly have in mind in this paper (see Sec. ~\ref{sec:Ising}).

Let us now specialize to a 2-level system coupled with a bath. If
interaction is weak enough, the two effective Hamiltonians are
slightly different. For an instance in Sec.~\ref{sec:Ising}, we will
consider the important case where the initial state of the bath is a
thermal equilibrium state $\rho_B=\exp(-\beta H)/Z(\beta)$
[$Z(\beta)=\Tr[\exp(-\beta H)]$ is the partition function,
$T=\beta^{-1}$ is the temperature of the bath]. In such a case $\mid
F_{\rho_B}(V_1,V_2)\mid^2$ will become the Loschmidt echo for
$\beta^{-1}=0$ (for non degenerate ground state) and  it coincides
with the operator fidelity defined in \cite{XWang} for $\beta=0.$

\section{Metrics over manifolds of unitaries\label{sec:unitary}}
In this section we discuss operator fidelity from a rather general mathematical point view.
Let $\cal H$ be a quantum state-space and a $\{ U_\lambda\}\subset {\cal U}({\cal H})$ a smooth family of unitaries over $\cal H$
parametrized by elements of a manifold $\cal M$.
Given a state $\rho\in {\cal S}({\cal H})=\{\rho\in {\cal L}({\cal H})\,|\,\rho\ge 0,\, \rm{tr}\rho=1\}$ one can define the following
hermitean product over ${\cal L}({\cal H})$
\begin{equation}
\langle X, Y\rangle_\rho:=\text{Tr}( \rho X^\dagger Y)
\label{scalar}
\end{equation}
If $\rho>0$ then (\ref{scalar}) is non-degenerate and $\|X\|_\rho:=\sqrt{\langle X, X\rangle_\rho}$
defines a norm over ${\cal L}({\cal H}).$ In general, if $\rm{ker}\rho\neq \{0\},$ $\|\bullet\|_\rho$ is just a semi-norm
(if the range of $X$ is included in $({\rm{supp}}\rho)^\perp$ then $\|X\|_\rho=0$)
Notice that i) $ \|X\|_\rho \le \|X\|$ and ii) unitaries $U$ are normalized i.e., $\langle U, U\rangle_\rho=1.$

{\bf{Definition}} The $\rho$-fidelity of  the operators $X$ and $Y$
is given by
\begin{equation}
{\cal F}_\rho (X,Y):= |\langle X, Y\rangle_\rho|
\label{op_fid}
\end{equation}
It is immediate yet important to realize that for non full-rank states $\rho$, having two unitaries with $\rho$-fidelity one does not imply their equality
(up to a phase). Indeed,
from the Cauchy-Schwarz inequality on has ${\cal F}_\rho (X,Y)\le \|X\|_\rho  \|Y\|_\rho$, in particular for unitaries $U$ and $W,$
${\cal F}_\rho(U,W)\le 1.$ One has that ${\cal F}_\rho(U,W)=1\Leftrightarrow U|_{\rm{supp}\rho}=W|_{\rm{supp}\rho}$
where supp$\rho:=(\rm{ker}\rho)^\perp.$

In order to unveil the operational meaning of the above definition  it is useful
to recall the following fact. If $|\Psi_\rho\rangle\in{\cal H}\otimes{\cal H}$ is a purification of $\rho=\sum_i p_i |i\rangle\langle i|$ i.e.,
$|\Psi_\rho\rangle=\sum_i \sqrt{p_i} |i\rangle\otimes|i\rangle$ then $\langle X, Y\rangle_\rho=\langle\psi_\rho|(X^\dagger\otimes\openone)(Y\otimes\openone)|\Psi_\rho\rangle.$
The operator scalar product (\ref{scalar}) can be seen as a scalar product of suitable quantum states of a bigger system.
This simple remark shows that operator fidelity (\ref{op_fid}) quantifies the degree of statistical distinguishability between the two states
$$|\Psi(A)\rangle:= (A\otimes\openone)|\Psi_\rho\rangle \,(A=X,Y).$$ When $U$ and $W$ denote unitary transformations
the quantity (\ref{op_fid}) has an interpretation as visibility strength in properly designed interferometric experiments \cite{oi}.
Yet another kind of operational relevance of the operator fidelities \ref{op_fid}
in the context of decoherence will be discussed in ~\ref{sec:model}.

Now, following the differential-geometric spirit of \cite{Zanardi2},
we are going to consider the operator fidelity (\ref{op_fid})
between infinitesimally different unitaries. The leading term in the
expansion of (\ref{op_fid}) will define a quadratic form over the
tangent space of the manifold ${\cal U}({\cal H}).$ For full rank
$\rho$ that quadratic form is  a {\em metric}.
 The following proposition shows that and its proof it is just a direct calculation analogous to the one
performed at the state-space level \cite{Zanardi2}.

{\bf{Proposition}}
Let $\{ U_\lambda\}\subset {\cal U}({\cal H})$ be a smooth family of unitaries over $\cal H$
parametrized by elements $\lambda$ of a manifold $\cal M$. One finds
\begin{equation}
{\cal F}_\rho(U_\lambda, U_{\lambda+\delta\lambda})=1
-\frac{\delta\lambda^2}{2}  \chi_\rho(\lambda)
\end{equation}
where, if $U^\prime=\partial U/\partial\lambda$ one has
\begin{equation}
\chi_\rho(\lambda):=\langle U^\prime, U^\prime\rangle_\rho -|\langle U^\prime, U\rangle_\rho|^2
\label{eq:chi}
\end{equation}
The quantity above will be referred to as operator fidelity
susceptibility (OFS). In the following we will be discussing three
different situations fitting in the framework we have set. The first
example ({\bf Ex.~0}) will show in which precise sense the operator
approach here under investigation includes the ground-state fidelity
one. In {\bf Ex.~1} we notice that when the family of unitaries is a
one-parameter group generated by an Hamiltonian $H$ then the metric
is nothing but the variance of $H$. Finally, in {\bf Ex.~2} the most
important case of a fixed-time family of evolutions generated by a
parametric family of Hamiltonians is analyzed.

{\bf{ Example 0}} Let $\{H_\lambda\}$ a family of {\em
isodegenerate} quantum Hamiltonians with spectral resolutions
$$H(\lambda)=\sum_n E_n(\lambda) |\psi_n(\lambda)\rangle\langle\psi_
n(\lambda)|.$$
 The { adiabatic intertwiners} are unitaries such
that $O(\lambda,\lambda_0)
|\psi_n(\lambda_0)\rangle=|\psi_n(\lambda)\rangle \forall n.$ Let
$\rho=|\psi_0(\lambda_0)\rangle\langle \psi_0(\lambda_0)|$
($|\psi_0(\lambda_0)\rangle$=Ground State of $H(\lambda_0)$) and
$U(\lambda)= O(\lambda,\lambda_0)$ then
\begin{eqnarray}
& & \langle U(\lambda),
U(\lambda^\prime)\rangle_\rho=\langle\psi_0(\lambda_0)|O^\dagger(\lambda,\lambda_0)O(\lambda^\prime,\lambda_0)|\psi_0(\lambda_0)\rangle=
\nonumber\\& &
\langle\psi_0(\lambda), \psi_0(\lambda^\prime)\rangle.
\end{eqnarray}
{ This shows that the ground state fidelity is a particular instance of the setting we are discussing}.

{\bf{Example 1}}
Let $\{U_t=e^{-it H}\}_{t}$ and $\rho=|\phi\rangle\langle\phi|.$ Then $U^\prime= - i H U(t)$ whence
$\chi_\rho(t)= \langle\phi| H^2\phi|\rangle -|\langle\phi| H|\phi\rangle|^2$

{\bf{Example 2}}
$\{U_\lambda=e^{-it H_\lambda}\}_\lambda$ and $\rho=\openone/d$ ($d:=\rm{dim}{\cal H}).$ From time-dependent perturbation theory one obtains
\begin{equation}
U^\prime=-i U_\lambda \int_0^t d\tau e^{i\tau H_\lambda} H^\prime e^{-i\tau H_\lambda}
\end{equation}
Introducing the superoperator ${ L}_H:=[H,\bullet]$ over ${\cal L}({\cal H})$
one sees that, if
$|n\rangle$'s and the $E_n$'s denote the eigenvectors and eigenvalues of $H(\lambda),$
 $E_n-E_m$ and $\hat\Psi_{n,m}:=\sqrt{d} |n\rangle\langle m|$ are the  eigenvalues and (normalized) eigenoperators of $L_H$, respectively.
One has
$\langle U^\prime, U^\prime\rangle=\langle \int_0^td\tau e^{i\tau L_H}(H^\prime), \int_0^td\sigma e^{i\sigma L_H}(H^\prime)\rangle.$
If $P=\sum_n|\hat{\Psi}_{n,n}\rangle\langle\hat{\Psi}_{n,n}|$ is the projection over the kernel of
$L_H$ and $Q=1-P$ one finds $ e^{i\tau L_H}= Q e^{i\tau L_H}Q +P,$ therefore the above expression becomes
\begin{eqnarray}
& &\langle U^\prime, U^\prime\rangle= \langle Q \frac{e^{i\tau L_H}-1}{L_H}Q(H^\prime),Q \frac{e^{i\tau L_H}-1}{L_H}Q(H^\prime)\rangle +\nonumber \\
& &t^2 \langle P(H^\prime),P(H^\prime)\rangle
=\| F_t Q (H^\prime)\|^2+t^2\|P(H^\prime)\|^2
\label{abs}
\end{eqnarray}
(notice that $L_H$ is invertible, by definition, on the range of $Q$) where  $F_t(x):=\sin(xt/2)/(x/2).$

The first term  in the last line of (\ref{abs}) can be rewritten as
$ \langle H^\prime|Q F_t^2(L_H) Q|H^\prime\rangle. $
Similarly the  term $\langle U,
U^\prime\rangle$  in the metric can be written as $\langle
1,\int_0^t d\tau e^{i\tau L_H}(H^\prime)\rangle=
\int_0^td\tau\langle e^{-\tau L_H}(1), H^\prime\rangle=t \langle 1,
H^\prime\rangle =d^{-1} {\rm{Tr}} H^\prime.$ Putting all together
and expanding over the basis $|\hat\Psi_{n,n}\rangle$ of
$L_H$-eigenoperators one finds the explicit result
\begin{eqnarray}
\chi_\rho(\lambda)
&=&\frac{1}{d} \sum_{n\neq m}|\frac{ \langle n|H^\prime|m\rangle}{E_n-E_m}|^2 F_t^2(E_n-E_m)\\
 &+&t^2(\frac{1}{d}\sum_n |\langle n|H^\prime|n\rangle|^2 - \frac{1}{d^{2}}|{\rm{Tr}}H^\prime|^2)
\label{G}
\end{eqnarray}
The key
property  here is that $$\lim_{t\to\infty} t^{-1}
F^2_t(x)=2\pi\delta(x).$$
This asymptotic delta function is
responsible for the large contribution to (\ref{G})given by small $E_n-E_m$
This shows that  {\em all} (quasi) level crossings in the spectrum of $H$
might lead to a analyticity breakdown in $\chi_\rho.$ This has to be contrasted  with
the GS fidelity studied e.g., in \cite{Zanardi} where just (quasi) level crossings in the GS
play a role. In this  sense it is clear why the OFS (\ref{G}) is a more powerful tool
than the corresponding state-space analogue.

An important generalization of (\ref{G}) can be obtained by
considering $\rho$ {\em commuting} with $H.$ In this case one
obtains an expression analogous to (\ref{G}) with the diagonal
elements $\rho_{n,n}=\langle n|\rho|n\rangle$ suitably inserted (see
Sec.~\ref{sec:analysis}).

\subsection{Splitting $\chi_\rho$\label{sec:analysis}}
Now we derive an alternative form for the OFS (\ref{eq:chi}) for the case of {\em Ex. 2} discussed above.
This form will make even more manifest the
different contributions to operator suceptibility arising from eigenvalue and eigenvector
variation with the control parameter $\lambda.$
An unitary operator can be written in diagonal form
\begin{equation}
  U(\lambda)=\sum_iu_i(\lambda)P_i(\lambda).
\end{equation}
where the $u_i$'s are  the eigenvalues of $U$ satisfying $|u_i|=1$
and $P_i=\matBasis{\psi_i}{\psi_i}$ is an one-dimensional projective
operator. Both eigenvalues and eigenstates are parameter dependent.
What we would like to do first is to separate the contributions of
these two kinds of parameter dependence to the susceptibility
$\chi_\rho(\lambda)$. In Ref.~\cite{Zanardi4}, the authors succeed
in telling apart the classical and quantum contribution to the Bures
metric with consideration of these two kinds of dependence.

The differential of this unitary operator can be divided into two
parts
\begin{align}
  \label{eq:pt_U}
  \nonumber
  \partial_\lambda U(\lambda)&=D_1(\lambda)+D_2(\lambda),\\
  D_1(\lambda)&=\sum_i(\partial_\lambda u_i)P_i,\quad D_2(\lambda)=\sum_i u_i(\partial_\lambda P_i),
\end{align}
where $\partial_\lambda
P_{i}=\matBasis{\psi_i}{\partial_\lambda
  \psi_i}+\matBasis{\partial_\lambda
  \psi_i}{\psi_i}$. Next, we assume that the
density matrix $\rho$ can be simultaneously diagonalized with $U$, so
it can be written as the form
\begin{equation}
  \rho=\sum_i \rho_{ii}P_i.
\end{equation}
The assumption $[H,\,\rho]=0$ is motivated by the important example where both time evolution unitary
operators and density operators considered are generated by the same
Hamiltonian i.e., $\rho$ is the Gibbs thermal state associated with $H.$

Since
\begin{equation}
  \average{P_j \partial_\lambda P_i}_\rho
  =\delta_{ij}\rho_{jj}(\innerprod{\psi_j}{\partial_\lambda \psi_i}+\innerprod{\partial_\lambda \psi_i}{\psi_j})
  =0,
\end{equation}
after substituting Eq.~(\ref{eq:pt_U}) into Eq.~(\ref{eq:chi}), we
obtain
\begin{align}
  \label{eq:chi_D}
  \chi_\rho(\lambda)&=\chi_\rho^{(1)}(\lambda)+\chi_\rho^{(2)}(\lambda),\\
  \nonumber
  \label{eq:chi_single}
  \chi_\rho^{(1)}(\lambda)&=\average{D_1^\dag D_1}_\rho-|\average{U^\dag D_1}_\rho|^2,\\
  \chi_\rho^{(2)}(\lambda)&=\average{D_2^\dag D_2}_\rho.
\end{align}
Here we separate the contributions of $D_1$ and $D_2$. To make them
explicit, we consider the time evolution operator
$U(\lambda,t)=e^{-itH(\lambda)}$, and assume that the eigenstates
changing with $\lambda$ can be connected through a smooth unitary
transformation $\mathcal{S}(\lambda)$ which is time-independent and satisfies
$\mathcal{S}(\lambda)H(\lambda)\mathcal{S}(\lambda)^\dag=H_d(\lambda)$,
where $H_d$ is a diagonal Hamiltonian in the fixed $\lambda$-independent basis.
Therefore, the unitary operators
considered can be written
\begin{align}
  U(\lambda,t)= \mathcal{S}^\dag(\lambda)
  U_d(\lambda,t)\mathcal{S}(\lambda),
\end{align}
with $U_d=\exp(-it H_d)$. After differentiating with respect to
$\lambda$, we get
\begin{align}
  \nonumber
  D_1(\lambda,t)&=-it \mathcal{S}^\dag(\lambda) [\partial_\lambda H_d(\lambda)]\mathcal{S}(\lambda) U(\lambda,t),\\
  D_2(\lambda,t)&=[\mathcal{A}(\lambda),U(\lambda,t)],
\end{align}
where
\begin{align}
  \label{eq:A}
  \mathcal{A}(\lambda)&=(\partial_\lambda \mathcal{S}(\lambda))^\dag
  \mathcal{S}(\lambda).
\end{align}
Substituting Eq.~(\ref{eq:A}) into Eq.~(\ref{eq:chi_single}) leads to
\begin{align}
  \label{eq:chi1}
  \chi_\rho^{(1)}(\lambda,t)=t^2[\average{(\partial_\lambda
    H_d)^2}_\rho-\average{\partial_\lambda H_d}_\rho^2].
\end{align}
This can be considered as the fluctuation of the quantity
$\partial_\lambda H_d(\lambda)$ under the state $\rho$, with an extra
factor $t^2$.

The second term of Eq.~(\ref{eq:chi_D}) will be
\begin{align}
  \label{eq:chi2}
  \chi_\rho^{(2)}(\lambda,t)=\left \langle
    [U^\dag(\lambda,t),\mathcal{A}^\dag(\lambda)][\mathcal{A}(\lambda),U(\lambda,t)
    ]\right \rangle_\rho.
\end{align}
Since $U$ and $\rho$ can be diagonalized simultaneously, the above expression can be further rewritten  as
$  \chi_\rho^{(2)}(\lambda,t)=2 \sum_{mn}\rho_{nn}|\mathcal{A}_{mn}|^2\left(1-\cos[(E_n-E_m)t]\right).
$
This form explicitly shows that the time-dependent terms in $\chi_\rho^{(2)}(\lambda,t)$ are circular functions.

Now we would like to make a few general comments about
Eqs.~(\ref{eq:chi1}) and (\ref{eq:chi2}):
\begin{itemize}
\item[i)]
They correspond one to one
to the two terms in Eq. (\ref{G}) [with suitably inserted $\rho_{n,n}$]. In particular this remark shows that
${\cal A }_{n,m}= (E_n-E_m)^{-1} \langle n|H^\prime|m\rangle.$
Notice that $\cal A$ is nothing but the infinitesimal generator of the adiabatic intertwiner mentioned in Sec ~\ref{sec:unitary}.

\item[ii)] $\chi_\rho^{(i)} (i=1,2)$  separate apart the contributions of
the  eigenvalue variations from those of the eigenstates. In
Ref.~\cite{Zanardi4}, a similar distinction was made for the Bures
metric on thermal state manifolds (the corresponding terms  there
were named the classical and the quantum, respectively).
\item[iii)] Eqs ~(\ref{eq:chi1}) and (\ref{eq:chi2})
have quite different  forms of time dependence. $\chi_\rho^{(1)}$
gives $t^2$ contributions explicitly, while the second term consists
of circular functions with finite period of criticality.
When $t$ is large, if $\chi_\rho^{(1)}$ is not vanishing, it dominates the
OFS  and therefore the decay behavior of the operator fidelity.
\item[iv)]
If $\rho$ is pure [implying, since we have assumed the form $\rho=\sum_i \rho_{ii}P_i$, that the
initial state is an eigenstate of the Hamiltonian]  $\chi_\rho^{(1)}$
vanishes. Moreover if  $\mathcal{A}(\lambda)$ commutes with $U(\lambda,t)$, and therefore the eigenvectors of $U(\lambda)$ are
$\lambda$-independent, it is obvious from  (\ref{eq:chi2}) that $\chi_\rho^{(2)}$ vanishes.
\end{itemize}

\section{The transverse field Ising model\label{sec:Ising}}
In this section, we apply the general formalism developed so far to
the concrete case of the transverse field Ising mode. Before doing
that it is useful to discuss a bit  generally models having a
factorization structure.

\subsection{Factorizable models\label{sec:fact}}
In order to obtain the exact solution of $\chi_\rho(\lambda,t)$, we
analyze the cases where the unitary operator $U(\lambda,t)$ and thermal
state $\rho(\lambda,\beta)$ have the same factorization structure.
This assumption hold true, for example, in cases where they are both generated by
the same Hamiltonian $H(\lambda)$. These two quantities can be
then written in the composite space
\begin{eqnarray}
  \nonumber
  U(\lambda)&=&\bigotimes_{k=0}^M U_k(\lambda),\\
  \rho(\lambda)&=&\bigotimes_{k=0}^M \rho_k(\lambda),
\end{eqnarray}
where $U_k$ is still an unitary operator and $\rho_k$ is still a
density operator, corresponding to the $k$-th subspace. Note that
\begin{equation}
  \partial_\lambda U(\lambda) =
  \sum_{l=0}^M[\bigotimes_{k=0}^{l-1}U_k(\lambda)]\otimes\partial_ \lambda
  U_l(\lambda)\otimes[\bigotimes_{k=l+1}^MU_k(\lambda)].
\end{equation}
Since  $\Tr(A\otimes B)=\Tr (A) \Tr (B)$ and $\Tr(\rho_l)=1$, the
first and second terms of Eq.~(\ref{eq:chi}) can be expressed
respectively as
\begin{align}
  \label{eq:factorize_tr}
  \nonumber
  \Tr[\rho(\partial_\lambda U)^\dag(\partial_\lambda U)] &=\sum_{l}\Tr_l[\rho_l(\partial_\lambda U_l)^\dag(\partial_\lambda U_l)] +\sum_{l\neq l^\prime} A_{ll^\prime}\\
  \mid\Tr[\rho U^\dag(\partial_\lambda
  U)]\mid^2&=\sum_{l}\mid\Tr_l[\rho_l U_l^\dag(\partial_\lambda U_l)]\mid^2+\sum_{l\neq
    l^\prime}A_{ll^\prime}
\end{align}
where $A_{ll^\prime}=\Tr _l[\rho_l U_l^\dag(\partial_\lambda
U_l)]\Tr_{l^\prime}[\rho_{l^\prime} U_{l^\prime}^\dag(\partial_\lambda
U_{l^\prime})]$, and the subscript $l$ means the $l$-th subspace.

Thus, the OFS $\chi_\rho$ for the factorized unitary
operator is
\begin{eqnarray}
  \label{eq:G_dcmp}
  \chi_\rho(\lambda)&=&\sum_l \chi_{\rho,l}(\lambda)\\
  \label{eq:G_l}
  \nonumber
  \chi_{\rho,l}(\lambda)&=&\Tr_l[\rho_l(\partial_\lambda U_l)^\dag(\partial_\lambda U_l)]\\
  &&-\mid \Tr_l[\rho_l U_l^\dag(\partial_\lambda U_l)]\mid^2
\end{eqnarray}

\subsection{Ising Hamiltoninan\label{subsec:Ising}}
We now  consider transverse field Ising model. The Hamiltonian is
given by
\begin{equation}
  H(\lambda)=-J\sum_{l=-M}^{M}(\sigma_l^x \sigma_{l+1}^x + \lambda\sigma_l^z),
\end{equation}
where $J$ is an exchange constant hereafter assumed to be the unity, and
$\lambda$ is  the transverse field strength. This model can be
calculated analytically by using the Jordan-Wigner transformation
\cite{Lieb}
\begin{eqnarray}
  \nonumber
  \sigma _{i}^{x} &=&\prod_{j<i}(1-2c_{j}^{\dag }c_{j})(c_{i}+c_{i}^\dag), \\
  \sigma _{i}^{z} &=&1-2c_{i}^{\dag }c_{i},
\end{eqnarray}
which maps the spins to fermions. After the Fourier transformations,
the Hamiltonian in the momentum space is
\begin{align}\label{eq:H2}
  \nonumber
  H(\lambda)=&-\sum_{k=-M}^{M}[\cos (\frac{2\pi k}{N})-\lambda ](d_{k}^{\dag}d_{k}+d_{-k}^{\dag }d_{-k}-1)\\
  &+i\sin (\frac{2\pi k}{N})(d_{k}^{\dag}d_{-k}^{\dag }-d_{-k}d_{k}),
\end{align}
where $N=2M+1$. This Hamiltonian can be exactly solved by a Bogoliubov
transformation \cite{Sachdev}. However, introducing a set of
pseudo-spin operators is more convenient here. Since $n_k-n_{-k}$
($n_k=d_k^\dag d_k$) commutes  with every term of the
Hamiltonian~(\ref{eq:H2}), the pseudo Pauli operators can be defined
by \cite{Anderson}
\begin{eqnarray}
  \nonumber
  \varsigma_{kx}&=&d_k^\dag d_{-k}^\dag + d_{-k}d_k,\\
  \nonumber
  \varsigma_{ky}&=&-i(d_k^\dag d_{-k}^\dag - d_{-k}d_k),\\
  \varsigma_{kz}&=&d_k^\dag d_{k}+ d_{-k}^\dag d_{-k}-1.
\end{eqnarray}
These give the Pauli matrix in the $n_k-n_{-k}=0$ subspace, and become
zero matrix in the $n_k-n_{-k}=\pm 1$ subspaces. Therefore $\varsigma_{0x}$,
$\varsigma_{0y}$, $\varsigma_{0z}$ are just the standard Pauli matrices
$\sigma_{0x}$, $\sigma_{0y}$, and $\sigma_{0z}$.

With these operators, the Hamiltonian can be written as
\begin{align}
  \nonumber
  H(\lambda)&=\sum_{k=1}^{M}\mathcal{S}_k^\dag (\lambda) H_{k,d}(\lambda) \mathcal{S}_k (\lambda)+(\lambda-1)\varsigma _{0z},\\
  \nonumber
  H_{k,d}&=\Omega_k \varsigma_{kz},\\
  \mathcal{S}_k(\lambda)&=\exp(-i\frac{\theta _{k}}{2}\varsigma_{kx}),
\end{align}
where
\begin{align}
  \nonumber
  \Omega_k &= -2 \sqrt{[\lambda-\cos(2\pi k / N)]^2+\sin^2(2 \pi k/N)},\\
  \theta_k &= \arcsin[\frac{2\sin(2 \pi k / N)}{\Omega_k}].
\end{align}
We consider the fidelity of two time evolution operators
$U(\lambda)=\exp(-itH(\lambda))$ and $U(\lambda+\delta\lambda)$ with
thermal state $\rho=\exp(-\beta H(\lambda))/Z(\beta,\lambda)$, where
$Z(\beta,\lambda)=\Tr [\exp(-\beta H(\lambda))]$ is the partition
function, and $\beta=1/T$ is the inverse temperature. The operators
$U$ and $\rho$ can be written in the factorized form
\begin{eqnarray}
  \nonumber
  U&=&e^{-it(\lambda-1) \sigma_{0z}}\otimes [\bigotimes_{k=1}^M U_k],\\
  \rho &=&\frac{e^{\beta (\lambda-1)\sigma _{0z}}}{Z_{0}}\otimes [\bigotimes_{k=1}^{M}\frac{\varrho _{k}}{Z_{k}}],
\end{eqnarray}
where
\begin{align}
  \nonumber
  U_k&=\mathcal{S}^\dag_k(\lambda) e^{-it \Omega_k \varsigma_{kz}}\mathcal{S}_k(\lambda)=e^{-it \Omega_k \varsigma_{kn}},\\
  \nonumber
  \rho_k&=\mathcal{S}_k^\dag (\lambda)e^{-\beta \Omega_k \varsigma_{kz}}\mathcal{S}_k(\lambda)=e^{-it \Omega_k \varsigma_{kn}},\\
  \varsigma_{kn}&=\varsigma_{ky}\sin\theta_k+\varsigma_{kz}\cos\theta_k,
\end{align}
and the partition functions are
\begin{align}
  Z_{0}=2\cosh (\beta (\lambda-1)),\quad Z_{k}=2[1+\cosh (\beta \Omega
  _{k})].
\end{align}
For $k>0$, after substituting $H_{k,d}= \Omega_k \varsigma_{kz}$ into
Eq.~(\ref{eq:chi1}), we have
\begin{align}
  \chi_{\rho,k}^{(1)}(\lambda,t)=4t^2\frac{\cos^2 \theta_k}{\cos(\beta
    \Omega_k)+1}.
\end{align}
To calculate $\chi_{\rho,k}^{(2)}$, we should first calculate the
related quantities $\mathcal{A}_k$ defined by Eq.~(\ref{eq:A}) and the
commutator
\begin{align}
  \nonumber
  \mathcal{A}_k&=i \frac{\theta_k^\prime}{2}\varsigma_{kx},\\
  [\mathcal{A}_k,U_k]&=i\theta_k^\prime
  \sin(t\Omega_k)(\varsigma_{kz}\sin
  \theta_k-\varsigma_{ky}\cos\theta_k),
\end{align}
where $\theta_k^\prime$ denotes $\partial_{\lambda}\theta_k$ for
convenience. Substituting them into Eq.~(\ref{eq:chi2}), we obtain
\begin{align}
  \chi_{\rho,k}^{(2)}&= 4\frac{\cosh (\beta \Omega _{k})}{\cosh (\beta
    \Omega _{k})+1}\frac{\sin ^{2}\theta _{k}\sin^2
    (t\Omega_{k})}{\Omega _{k}^{2}}.
\end{align}
For $k=0$ subspace, the contribution is
\begin{align}
 \chi_{\rho,0}^{(1)}= t^2[1-\tanh ^2(\beta (\lambda-1))], \quad
 \chi_{\rho,0}^{(2)}&=0.
\end{align}
Thus, we get the OFS
\begin{align}
\nonumber
\chi_\rho^{(1)}&=t^2[1-\tanh ^2(\beta (\lambda-1))]+4\sum_{k=1}^{M}t^2\frac{\cos^2 \theta_k}{\cos(\beta\Omega_k)+1},\\
\chi_\rho^{(2)}&=4\sum_{k=1}^M\frac{\cosh (\beta \Omega _{k})}{\cosh (\beta
  \Omega _{k})+1}\frac{\sin ^{2}\theta _{k}\sin^2
  (t\Omega_{k})}{\Omega _{k}^{2}}.
\end{align}
Notice that these expressions, for the XY model, can be also obtained directly from Eq. (\ref{G}).
For a given $\lambda$, we can consider the time average OFS which is given  by
\begin{align}
    \overline{\chi_\rho}(\lambda)=\lim_{T\rightarrow \infty}\frac{1}{T}\int_0^T dt \chi_\rho(\lambda,t).
\end{align}
Obviously $\overline{\chi_\rho^{(1)}}(\lambda)$ diverges, since $\chi_\rho^{(1)}(\lambda)$ is proportional to $t^2$. The time average
of the circular function contributions $\chi_\rho^{(2)}$ is
\begin{align}
\overline{\chi_\rho^{(2)}}(\lambda)=2\sum_{k=1}^M\frac{\cosh (\beta \Omega _{k})}{\cosh (\beta
  \Omega _{k})+1}\frac{\sin ^{2}\theta _{k}}{\Omega _{k}^{2}}.
\end{align}
After rescaling $\chi_\rho \rightarrow \chi_\rho/N, 2\pi k/N
\rightarrow k$  and taking the thermodynamical limit, we get
\begin{align}
  \nonumber
  \chi_\rho^{(1)}(\lambda,t)&=\frac{2 t^2}{\pi}\int_{0}^{\pi}dk \frac{1}{\cosh (\beta \Omega _{k})+1}\cos ^{2}\theta_{k},\\
  \nonumber
  \chi_\rho^{(2)}(\lambda,t)&=\frac{2}{\pi}\int_{0}^{\pi}
  dk\frac{\cosh (\beta \Omega _{k})}{\cosh (\beta \Omega
    _{k})+1}\frac{\sin ^{2}\theta _{k}\sin^2 (t\Omega_{k})}{\Omega
    _{k}^{2}}.\\
\label{eq:chi2_Ising}
 \overline{\chi_\rho^{(2)}}(\lambda)& = \frac{1}{\pi}\int_{0}^{\pi} dk\frac{\cosh (\beta \Omega
    _{k})}{\cosh (\beta \Omega _{k})+1}\frac{\sin ^{2}\theta
    _{k}}{\Omega _{k}^{2}}.
\end{align}

\begin{figure}[htbp]
  \centering{\epsfig{scale=0.3,file=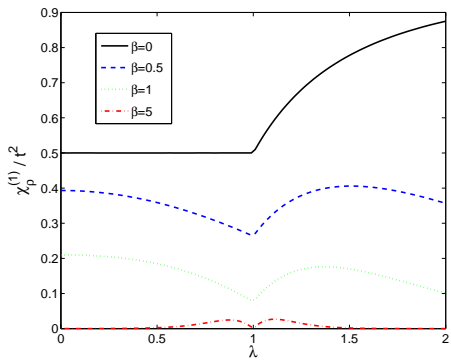}}
  \caption{First term of OFS  divided by $t^2$ as a
    function of $\lambda$ with different temperatures in the thermodynamical limit.}
  \label{Fig:1}
\end{figure}
\begin{figure}[htbp]
  \centering{\epsfig{scale=0.3,file=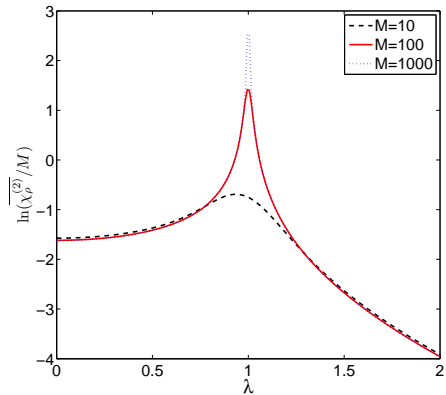}}
  \caption{Time average of the second term of OFS as a
    function of $\lambda$ for a given $\beta=1$ with different $M$.}
  \label{Fig:2}
\end{figure}

\begin{figure}[htbp]
  \centering{\epsfig{scale=0.4,file=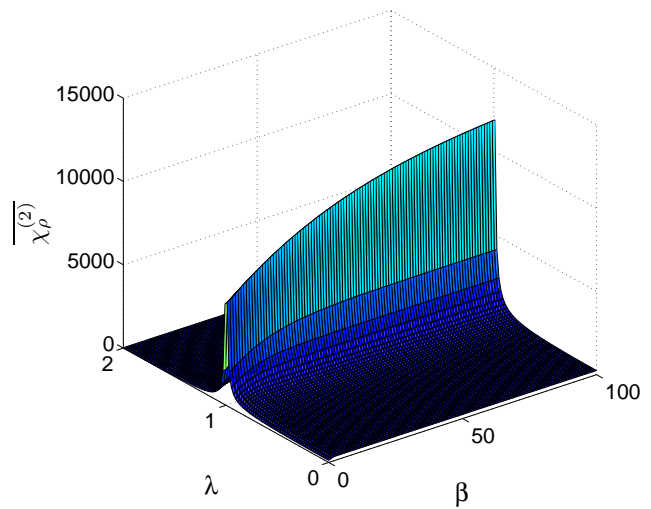}}
  \caption{Time average of the second term of OFS as a
    function of $\lambda$ and $\beta$ for a given $M=1000$.}
  \label{Fig:3}
\end{figure}

One finds that both the quantities ${\chi_\rho^{(1)}}(\lambda,t)$ and $\overline{\chi_\rho^{(2)}}(\lambda)$ have nontrivial behavior
at the critical point $\lambda=1$ and, therefore, can be used as  indicators of criticality.
At
$\lambda=1, $ $\chi_\rho^{(1)}$ has a  minimum (see
Fig.~\ref{Fig:1}),  while $\overline{\chi_\rho^{(2)}}$ has a maximum,
(see Fig.~\ref{Fig:3}).

Notice that $\chi_\rho^{(1)}$ vanishes only when the inverse
temperature $\beta=\infty$. When the temperature is not zero, for
long enough time, $\chi_\rho$ will be dominated by $\chi_\rho^{(1)}$
and therefore will have a minimum at $\lambda=1$. This has to be contrasted
with  the previous results for ground states in Ref. \cite{HTQuan} where is was argued that
the short time LE decay is enhanced at criticality (large $N$.)
For a
fixed $t$ and $\lambda$, as temperature increases, $\chi_\rho^{(1)}$
grows  while $\chi_\rho^{(2)}$ decays.

The time average of second term of OFS will diverge at the critical
points $\lambda=1$ under thermodynamical limit (see Fig.
\ref{Fig:2}). It is caused by the denominator $\Omega_k$ in
Eq.~(\ref{eq:chi2_Ising}), since it will vanish for those very small
$k$s at the critical point $\lambda=1$ when $N$ goes to infinity.
Thus, this divergence will be retained for all $\beta$ (see
Fig.~\ref{Fig:3}).

In principle, one obtains ${\chi_\rho^{(1)}}(\lambda,t)$ and
$\overline{\chi_\rho^{(2)}}(\lambda)$ from experiments in different
ways. If we use ${\chi_\rho^{(1)}}(\lambda,t)$ to investigate the
quantum critical point, the measurement of the time interval $t$ is
important. However, when ${\chi_\rho^{(1)}}(\lambda,t)$ vanishes, we
can use $\overline{\chi_\rho^{(2)}}(\lambda)$ to investigate quantum
criticality. This means we can get a $\chi_\rho^{(2)}(\lambda,t)$
for a random $t$ again and again , then evaluate the time average
value. By this scheme, we can avoid the demand of the exact
measurement of the time interval $t$ from the coupling instant to
the measuring instant.

\section{conclusion\label{sec:conclusion}}
To summarize, we have introduced the state-dependent operator
fidelity and its susceptibility $\chi_\rho$ to study  environment-
induced decoherence. By deriving general expressions for $\chi_\rho$
we identified two different contributions to
the susceptibility. These two terms have different physical origin and temporal
behavior. For the transverse field Ising model, we obtained an exact
expression for $\chi_\rho$ and showed that it has  nontrivial
behavior at the critical point both at zero and non-zero
temperature. Moreover, from $(\ref{G})$ it is clear
that the OFS depends crucially in the level spacing distribution of $H$. This leads to conjecture that OFS
might be an effective tool to investigate the transition to quantum chaos as well.
Finally, we believe this type of analysis is directlly relevant to
experiments aimed at using quantum probes to  detect QPTs.
\section{Acknowledgments}
We thank N. Toby Jacobson for a careful reading of the manuscript.
The work was supported by the Program for New Century Excellent
Talents in University (NCET), the NSFC with grant nos. 90503003, the
State Key Program for Basic Research of China with grant nos.
2006CB921206, the Specialized Research Fund for the Doctoral Program
of Higher Education with grant No.20050335087.

\end{document}